# Strange Metallic Transport in the Antiferromagnetic Regime of Electron Doped Cuprates

Tarapada Sarkar[1], Nicholas R Poniatowski[1], Joshua S. Higgins[1], P. R. Mandal[1], Mun K Chan[2], and Richard L Greene[1]

[1]*Maryland Quantum Materials Center and Department of Physics, University of Maryland, College Park, Maryland 20742, USA.*

[2]*Pulsed Field Facility, National High Magnetic Field Laboratory, Los Alamos National Laboratory, Los Alamos, NM 87545, USA.*

We report magnetoresistance and Hall Effect results for electron-doped films of the high-temperature superconductor $La_{2-x}Ce_xCuO_4$ (LCCO) for temperatures from 0.7 to 45 $K$ and magnetic fields up to 65 T. For x = 0.12 and 0.13, just below the Fermi surface reconstruction (FSR), the normal state in-plane resistivity exhibits a well-known upturn at low temperature. Our new results show that this resistivity upturn is eliminated at high magnetic field and the resistivity becomes linear-in-temperature from ~40 K down to 0.7 K. The magnitude of the linear coefficient scales with $T_c$ and doping, as found previously [1,2] for dopings above the FSR. In addition, the normal state Hall coefficient has an unconventional field dependence for temperatures below 50K. This anomalous transport data presents a new challenge to theory and suggests that the strange metal normal state is also present in the antiferromagnetic regime.

**Keywords:** Strongly correlated system, superconductors

# I. INTRODUCTION

An understanding of the copper oxide (cuprate) superconductors remains one of the major unsolved problems of condensed matter physics. It is generally agreed that to understand the origin of the high-Tc superconductivity one has to understand the nature of the low temperature normal state from which the superconductivity emerges. A key unexplained feature of the cuprates is the upturn in the normal state $CuO_2$ plane resistivity, $\rho(T)$ for $H > H_{c2}$, seen at low temperature for doping below the Fermi surface reconstruction (FSR) in electron-doped (n-doped) [3,4] and in the pseudogap phase for hole-doped (p-doped) [5-12]. Various explanations have been proposed, such as disorder, Kondo scattering, or loss of carriers [5-7, 10, 11], but there is no consensus on the origin of this upturn. For very low doping (far from the FSR) the normal state resistivity has an insulator-like behavior as the Mott-Hubbard doping (x = 0) is approached and there is no superconductivity at zero field. In this doping regime, the resistivity increases logarithmically as the temperature approaches zero.

In electron-doped cuprates doped just below the FSR -- the subject of this paper -- the ab-plane resistivity upturn has been shown experimentally to be linked to spin scattering [3,13,14]. This is supported by a theoretical model that suggests magnetic droplets can be formed in regions of the phase diagram that exhibit short range antiferromagnetic (AFM) order [15]. In this regime of the temperature-doping phase diagram, the normal state resistivity develops the well-known upturn. The resistivity tends to saturate as the temperature approaches zero showing that the normal state is a metal and not an insulator. Recently, a similar link between AFM and the resistivity upturn has been proposed for p-doped $La_{2-x}Sr_xCuO_4$ (LSCO) [12].

At high doping (beyond the dome of superconductivity) the n-doped cuprates have a $T^2$ resistivity at low temperatures [1,16] as found for a Fermi liquid (FL) metal. Between the end of the SC dome and the FSR doping a non-FL behavior is found where $\rho(T) \propto T$ [1,2,]. Similar behavior is found in p-doped cuprates [17,18]. This anomalous low temperature transport has been called the strange metal [3, 19] and has become the subject of intense study.

An important unexplained issue is the nature of the low temperature normal state ($H > H_{c2}$) in the absence of the upturn. Recent work on n-doped cuprates has claimed that the ground state in the upturn region is a Fermi liquid [20]. This conclusion was based on a subtraction of the estimated upturn resistivity from the total resistivity and finding a $T^2$ temperature dependence of the remaining resistivity. Our conclusions, based on direct experimental measurements, are quite different.

Here, we study the magnetic field dependence of the ab-plane resistivity as a function of temperature in the upturn region of the n-doped cuprates LCCO and PCCO, an experiment that has not previously been performed for any cuprate, to our knowledge. For dopings just below the FSR doping, we find new and unexpected behaviors of the electrical transport (namely, the resistivity and Hall Effect). The resistivity upturn can be completely suppressed above a certain magnetic field ($H^*$) and above $H^*$ the resistivity exhibits a strange metallic linear-in-T dependence from 0.7K to ~40K and an anomalous linear-in-H magnetoresistance for $H > H^*$ in the same temperature range. Further, the Hall coefficient has an unconventional field dependence for T < ~50K.

Our conclusion is that the strange metal behavior of the normal state at low temperature persists at doping both above and below the FSR in n-doped cuprates, i.e., for all doping where superconductivity exists in zero field. We will give a few qualitative comments about these results and their implications, but it is unambiguous that our new data represents another challenge for the theory of the cuprates.

## II. METHODS

The high field magnetic measurements were performed on La$_{2-x}$Ce$_x$CuO$_4$(LCCO) films for two dopings (x=0.13, 0.12) just below the Fermi surface reconstruction doping (x = 0.14). The films, of thickness about 150 nm, were grown using the pulsed laser deposition (PLD) technique on SrTiO$_3$ [100] substrates (5 × 5$mm^2$) at a temperature of 750$^\circ$ C utilizing a KrF excimer laser. The films were post-annealed at 600$^\circ$ C in an oxygen partial pressure of about 1×10$^{-5}$ torr for 30 minutes to remove the apical oxygen and induce superconductivity. The full width at half maximum of the peak in $d\rho_{xx}/dT$ of the films is within the range of 0.5 K, demonstrating the high quality of the samples. The LCCO targets have been prepared by the solid state reaction method using 99.99% pure La$_2$O$_3$, CeO$_2$, and CuO powders. The Bruker x-ray diffraction (XRD) of the films shows the c-axis-oriented epitaxial LCCO tetragonal phase. The thickness of the films has been determined by using cross-sectional scanning electron microscopy (SEM). The high-field 65 T measurement was performed by standard four-probe ac lock-in method at the National High Magnetic Field Laboratory (NHMFL) Pulsed Field Facility, Los Alamos National Laboratory and the 35 T dc field measurements were performed at the National High Magnetic Field Laboratory, Tallahassee DC field lab, Florida.

## III. RESULTS

Figures 1(a) and (b) show the measured in-plane magnetoresistivity, $\rho_{xx}(T, H)$, of the x=0.12 and 0.13 samples up to fields of 65 T for numerous temperatures between 0.7 K and 44 K. The temperature dependence of the magnetoresistivity, extracted by taking cuts of each curve in Figs. 1(a) and (b) at a fixed field, is plotted in Figs. 1(c) and (d) for several different values of the field. At low fields, the low-temperature normal state resistivity develops the well-known upturn [4-7] mentioned above. The temperature at which the resistivity reaches a minimum, T$_m$, decreases as the field is increased, and eventually vanishes at a field which we will label H*. At this field, the resistivity is linear in temperature, as one can see from the 65 T (red) curve in Fig. 1(a) and the 60 T (black) curve in Fig. 1(b).

The temperature dependence of the resistivity above H*, in particular whether it remains linear or crosses over to another power law, cannot be ascertained for LCCO from our current data since H* ≈ 65 T was the maximum field available for the measurement. However, by re-analyzing our previous measurements of another electron-doped cuprate, Pr$_{2-x}$Ce$_x$CuO$_4$ (PCCO), we can address this issue. Fig. 2 shows the temperature dependence of the magnetoresistivity of an optimally doped (x= 0.15) PCCO sample for fields up to 80 T taken from Ref. 21. For this sample, H* ≈ 55 T and as seen in Fig. 2(b) the resistivity remains linear for all measured fields above H*. Fitting these resistivity curves to $\rho_{xx}(T, H) = \rho_{xx}(0, H) + A(x)T$, we find that the coefficient of the T-linear resistivity A(x) increases slightly with field from 0.17 $\mu\Omega - cm/K$ at 60 T to 0.2 $\mu\Omega - cm/K$ at 80 T. This slight increase may be a consequence of the emergence of a quadratic-in-field contribution to the magnetoresistance at temperatures above T$_c$ which causes the magnitude of the magnetoresistivity to be larger at higher temperatures (see Fig. 3 of Ref. 2).

In Fig. 3(a), we fit the resistivity vs. temperature curves for the x = 0.12 LCCO sample at 65 T and the x = 0.13 LCCO sample at 60 T to the form $\rho_{xx}(T, H) = \rho_{xx}(0, H) + A(x)T$ where $\rho_{xx}(0, H)$ is the resistivity at zero temperature and the appropriate field. The slight low temperature deviation from linearity in the x = 0.12 doping is likely a sign that 65 T is less than H*, i.e., not quite sufficient to completely suppress the resistivity upturn.

The fitted slopes of the T-linear resistivity for the x = 0.13 and x=0.12 are plotted as a function of doping in Fig. 3(b), along with previously measured values of A(x) for overdoped LCCO samples taken from Ref. 2. Past work on overdoped LCCO has established that A(x) ~ 1/x and scales with the critical temperature for dopings above the FSR in the strange metal regime. Here, we find that A(x) for the underdoped x = 0.13 and x=0.12 sample falls on the same A(x) ~ 1/x curve as the overdoped samples and scales the same way with $T_c$. Thus, the linear-in-T resistivity reported here for underdoped samples appears to be of the same origin as that seen in overdoped samples.

In addition to our magnetoresistivity measurements, we report the high-field Hall coefficient as a function of temperature and magnetic field for LCCO, x = 0.12 and x = 0.13. It is known from our prior work on under doped samples [4] that the low-field (below 14T) Hall coefficient is peaked at a doping-independent temperature of order 10 K. Fig. 4 shows the Hall coefficient measured from 2 K to 80 K as a function of magnetic field up to 35T. As shown in Fig. 4b, as the magnetic field increases the low temperature peak in the Hall coefficient decreases and vanishes at high field, similar to how the resistivity minima vanishes with higher field (see Fig. 1). In Figs. 4a and c we see that above ~10K the magnitude of the Hall coefficient decreases with increasing field and tends to saturate at high field. This indicates that the resistivity minima and Hall coefficient peaks are interlinked, as was suggested in the previous reports [4, 22]. Fig. 5 shows $R_{xy}$ versus field at various temperatures from which $R_H(H)$ is found at different temperatures $R_H = R_{xy}(H)/H$.

## IV. DISCUSSION

The temperature dependence of the high field resistivity and Hall coefficient (normal state) reported here for $x < x_{FSR}$ LCCO is qualitatively similar to that of the $x > x_{FSR}$ LCCO. Therefore, it seems reasonable to speculate that the effect of strong magnetic fields is to move the location of the Fermi Surface Reconstruction (putative QCP) to lower Ce doping. This idea was proposed to explain some magnetic field effects in hole-doped cuprates [23,24]. However, our high field Hall coefficient at 2K, $R_H \approx 0.5 \times 10^{-10} (\Omega - m)/T$ for x=0.13 doping (see Fig. 4b) is an order of magnitude lower than the large hole pocket $R_H$ we found above the FSR [4]. Also, in this scenario, one would expect to find a large hole-like FS at high fields. But, quantum oscillation experiments of electron-doped cuprates with $x < x_{FSR}$ report low-frequency oscillations from the reconstructed small hole-like pocket, even at 60 T [21,25,26], in conflict with such an interpretation. The high field quantum oscillation frequency also gives a small Fermi surface pocket size that is the same as that measured in zero-field by ARPES [27]. Thus, it is unlikely that our findings can be thought of in terms of a shift in the position of the FSR .

Another possible explanation for our results is the magnetic field suppression of the in-plane, short range AFM, spin scattering that was proposed to be responsible for the resistivity upturn [13]. At a field of 50T, the Zeeman energy ($g\mu_B B$) is approximately 60 K which is roughly ~$2T_{min}$ ($T_{min}$ is estimated by extrapolating the Fig. 2(a) inset plot to zero field, which is roughly 20 K). This means that the Zeeman energy $g\mu_B B$ at 50T is approximately the same as the energy corresponds to $2T_{min}$ (at zero B). Thus, an external 50 T field could greatly suppress the spin scattering responsible for the resistivity upturn.

Next, we comment on the linear temperature dependence of the resistivity in LCCO and PCCO for $x < x_{FSR}$ that emerges at high fields. This low-temperature linear-in-T behavior is the hallmark of the strange metal state observed in both electron- and hole-doped cuprates [19]. In particular, such a state is observed in electron-doped cuprates for all dopings between the FSR doping and the end of the SC dome. After application of a large external field our results indicate that the underlying metallic ground state of $x < x_{FSR}$ electron-doped cuprates is also a strange metal (at least for dopings near the FSR). Another feature of this strange metal state is a linear-in-H magnetoresistance at low temperatures, as was found for $x > x_{FSR}$ LCCO films [2]. Although our present experiments on LCCO x = 0.13 did not go to high enough field to measure this, we note that higher field experiments on related n-doped cuprates did observe a linear-in-H magnetoresistance from 55-90 T for temperatures below 30K [28]. Moreover, the fact that the coefficient of the linear-in-T resistivity scales with doping in the same manner as $x > x_{FSR}$ samples (see Fig. 3) further suggests that the "hidden" strange metal ground state of $x < x_{FSR}$ samples is of the same origin as the strange metal state found on the $x > x_{FSR}$ side of the phase diagram.

Our conclusions are in stark contrast to some prior work, which argued the normal ground state of electron-doped cuprates was best described as a Fermi liquid [20]. However, we note that this prior work was done at zero magnetic field and relied on an uncertain subtraction of an estimated upturn resistivity, whereas our work here is the first direct measurement of the metallic ground state hidden underneath the resistivity upturn.

Our results indicate that a strange metallic ground state is present in the electron-doped cuprates for all dopings within the superconducting dome and is thus a universal feature of the electron-doped cuprates. Such a strange metal state is observed in hole doped LSCO [17], Bi2201[18] and Tl2201[29]for dopings above the pseudogap end point to the end of the SC dome. This is in stark contrast to many unconventional SC [30,31] where linear-in-T resistivity is observed only at a single, ostensibly quantum critical, doping. Consequently, this universality poses a challenge to many developing theories of strange metallic transport, in particular those which attribute the linear-in-T resistivity to quantum critical points. Further, our results demonstrate that strange metallic transport, whatever its origin, is largely insensitive to the geometry of the Fermi surface, in that it is observed on either side of the FSR where the Fermi surfaces vary significantly.

The Hall effect is another unexplained anomalous property of the cuprates. Many proposals have been made [for example see 18, 32-35 and references therein] but there is no consensus yet. Here, we discuss the Hall coefficient in our two underdoped LCCO films and show that the field and temperature dependence is inconsistent with conventional Boltzmann theory.

From the known fermiology [25,26,27] of underdoped electron-doped cuprates, consisting of a small hole-like pocket and a large electron-like pocket, we compare our Hall coefficient measurements to the standard two carrier Boltzmann transport model [36]. The components of the resistivity tensor are given by

$$\rho_{xy} = BR_H = \frac{1}{e}\frac{(n_h\mu_h^2 - n_e\,\mu_e^2) + \mu_h^2\mu_e^2 B^2(n_h - n_e)}{(n_h\mu_h + n_e\mu_e)^2 + \mu_h^2\,\mu_e^2\,B^2\,(n_h - n_e)^2}B \ \dots\dots\dots\dots\dots\dots(1)$$

$$\rho_{xx}(B) = \frac{1}{e}\frac{(n_h\mu_h + n_e\,\mu_e) + (n_e\,\mu_e\mu_h^2 + n_h\,\mu_h\mu_e^2)B^2}{(n_h\mu_h + n_e\mu_e)^2 + \mu_h^2\,\mu_e^2\,B^2\,(n_h - n_e)^2}\dots\dots\dots\dots\dots\dots(2)$$

where $n_h(\mu_h)$ and $n_e(\mu_e)$ are the carrier density (mobility) of electrons and holes respectively. To reduce the number of free parameters we estimate the carrier densities from the area of each pocket, as determined from the ARPES measurements of x = .15 NCCO [27]. The three-dimensional carrier density for each carrier type is then given by $n_{3D} = \frac{n_{2D}}{c} = \frac{1}{c}\frac{2}{(2\pi)^2}A_{FS}$ where $c = 12$ Å is the c-axis lattice constant and the Fermi surface area $A_{FS}$ is 1% (18%) of the Brillouin zone for the hole (electron) pocket at 20 K [27], yielding $n_h \approx 1.0x10^{26}m^{-3}$ and $n_e \approx 1.8x10^{27}m^{-3}$ .

We estimate the hole mobility from quantum oscillations (QO) measurements. The low-frequency oscillations corresponding to the hole-pocket onset at 40 T, so assuming that $\omega_c\tau{\sim}1$ at this field we estimate the hole mobility from $\omega_c\tau = \mu_h B = \mu_h 40 {\sim} 1$, which gives $\mu_h = 0.025\ T^{-1}$. To estimate the electron mobility, we note that the Hall resistivity changes sign (see Fig. 4), crossing zero at a field $B_c \approx 20\ T$ for x = .13 LCCO and $B_c \approx 30\ T$ for x = 0.15 NCCO (see Fig.6-[26]). Assuming the size of the Fermi surfaces and hole mobility of LCCO are similar to NCCO and comparing to Eq. 1, $B_c = \sqrt{\frac{(n_h\mu_h^2 - n_e\mu_e^2)}{\mu_h^2\mu_e^2\,(n_e - n_h)}}$, which allows us to estimate the electron mobility to be $\mu_e \approx 5.3 \times 10^{-3}\ T^{-1}$ for x = 0.13 LCCO and $\mu_e \approx 4.5 \times 10^{-3}\ T^{-1}$ for x = 0.15 NCCO.

We now use these four parameters $(n_h, \mu_h, n_e, \mu_e)$ to fit the measured Hall coefficient (Fig. 4) of LCCO using Eq. 1 and the magnetoresistance using Eq. 2. We find the fits to be inconsistent with conventional two carrier Boltzmann transport as argued below.

A simulated behavior of the normal state Hall coefficient as a function of field at 2K is shown in Fig.4(d) using these estimated four parameters. The behavior is very different than the measured normal state Hall coefficient below ~10K that is shown in Figs. 4a or 4c.

We compare the sign of $R_H$ in the high field limit where Eq. 1 give a Hall coefficient, $R_H = \rho_{xy}/H = \frac{1}{e}\frac{1}{(n_h - n_e)}$. This equation suggests $R_H$ should be negative since we estimated $n_e > n_h$. In contrast, we find that, at low temperatures (<10 K) and high field, $R_H$ is positive (see Fig-4a and Fig. 6 of [26]).

Our estimated high field limit Hall Coefficient, $/R_H/ = \frac{1}{e}\frac{1}{(n_h - n_e)}$, from our estimated values of $n_h(1x10^{26}m^{-3})$ and $n_e(1.8x10^{27}m^{-3})$ is ~$36x10^{-10}\Omega - m/$. This is an order of magnitude larger than our measured value of $R_H$ shown in Fig.4.

The high field magnetoresistance (see the high filed regime of refs -21, 28) is linear-in-H, which is also inconsistent with two-carrier transport where a quadratic field dependence magnetoresistance is expected in a conventional two-band model.

Thus, the normal state Hall coefficient and the normal state MR for x = 0.12, 0.13 doped films (just below the FSR at x = 0.14) exhibit an anomalous strange metal behavior.

A prior study on PCCO films also reported that the high field Hall coefficient does not fit with conventional two carrier Boltzmann transport for dopings just below the FSR [37].

A modified model with consideration of field dependent spin scattering might explain our data; however, the development of such a theoretical model is outside the scope of the present experimental work.

## V. CONCLUSIONS

We have performed low-temperature, ab-plane resistivity and Hall-effect measurements of the electron-doped cuprate $La_{2-x}Ce_xCuO_4$ for dopings x=0.12 and 0.13 and $Pr_{2-x}Ce_xCuO_4$ for doping x=0.15 (just below Fermi Surface Reconstruction) at high magnetic fields. These strong fields suppress the low-temperature resistivity upturn to reveal a linear-in-T resistivity whose magnitude scales with that of the low-temperature linear-in-T resistivity of LCCO for doping above the FSR. This result implies that the normal metal state hidden beneath the resistivity upturn is the same strange metallic state observed in overdoped samples. Consequently, these results suggest that the starnge metal normal state is a universal feature of electron-doped cuprates for dopings within the SC dome. Further, the Hall Effect cannot be explained by conventional Boltzmann transport theory and it appears to be another unexplained aspect of strange metal behavior. We hope that these results and the insights derived from them stimulate further theoretical work to identify the origin of this strange metallic state.

**Acknowledgements**

We thank Sankar Das Sarma, Johnpierre Paglione, Ian Hayes, and Pavel Volkov for many fruitful discussions. We also thank Eun Sang Choi for his DC high field user support at Tallahassee High Field lab. This work was supported by the NSF under grant no. DMR 1708334 and Maryland Quantum Materials Center (QMC). The National High Magnetic Field Laboratory was supported by the NSF Cooperative agreement nos. DMR-1157490 and DMR-1644779. M.K.C. acknowledges support by the Laboratory Directed Research and Development program of Los Alamos National Laboratory under project number 20180137ER for designing equipment for Pulsed-Field Experiments.

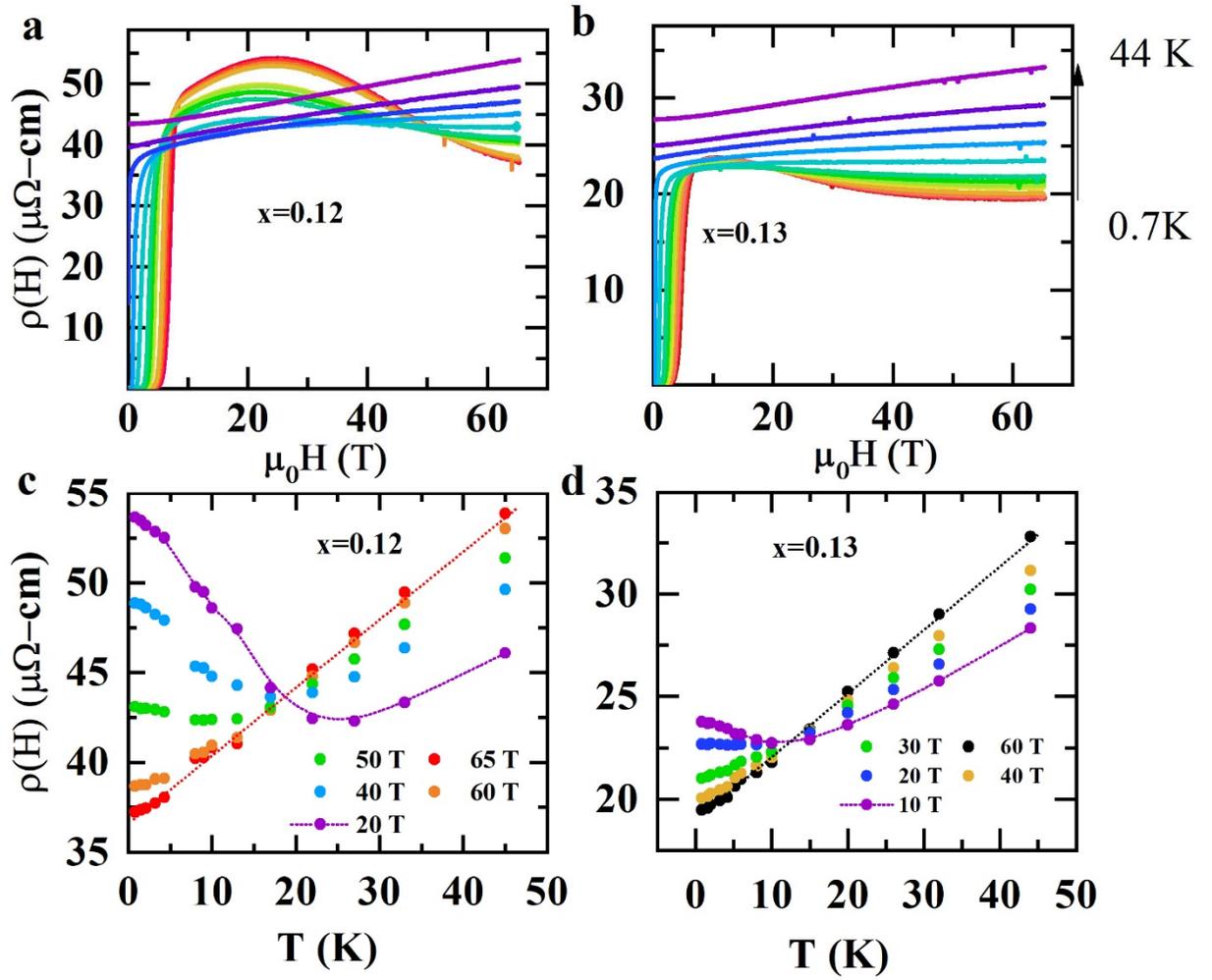

FIG. 1. Temperature dependent magnetoresistance ($\rho_{xx}(T,H)$ ) of LCCO for x=0.12(a) and x=0.13(b) ($H \perp ab$-plane). Black arrow indicates the increasing temperature direction from 0.7 K to 44 K. (measured in 65 T pulsed field). In (c) and (d) the resistivity vs temperature is found from the (a) and (b) data, respectively.

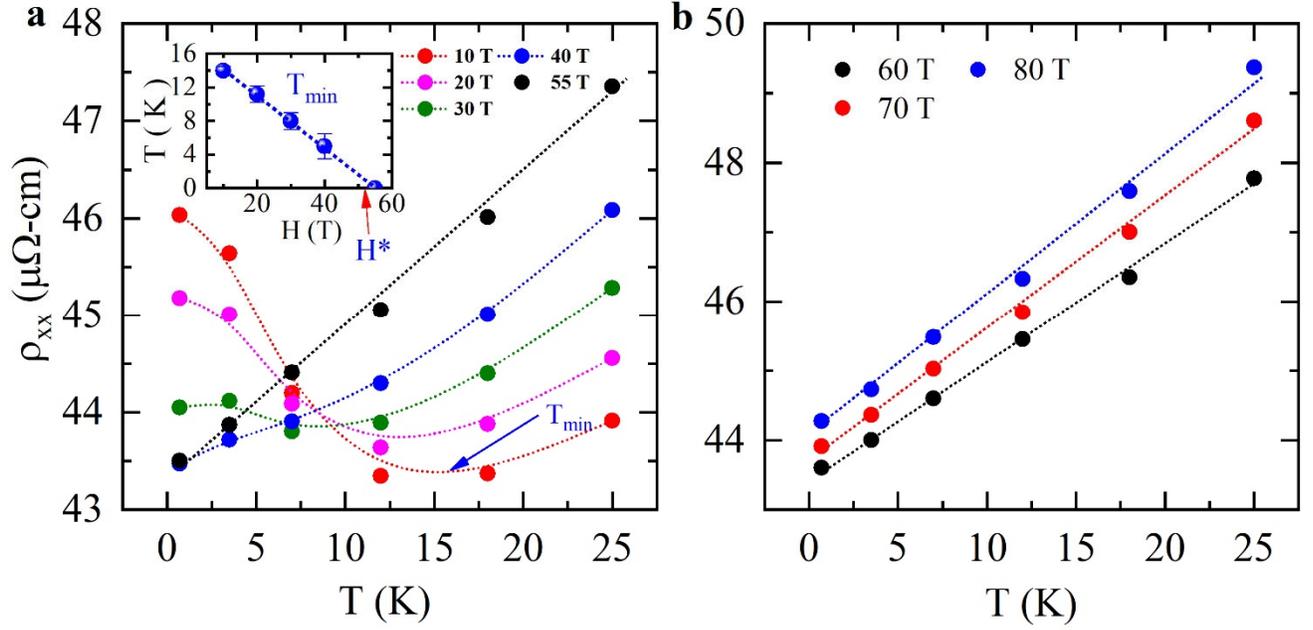

FIG. 2. (a) Transverse magnetoresistivity ($\rho_{xx}(T,H)$) vs temperature of $Pr_{2-x}Ce_xCuO_4$ for x=0.15 (data taken from Ref. [19]). The dotted lines are guides to the eye. Inset: the resistivity minima ($T_m$) vs field. The $T_m$ is determined by taking the derivative of the dotted lines and $H^*$ is the field where $T_m$ vanishes. (b) Resistivity vs T at higher fields. The dotted lines are a linear fit with $\rho_{xx}(T,H) = \rho_{xx}(0,H) + A(x)T$, with A(x) = .17 (60 T), .19 (70 T), and .2 (80 T) micro-ohm-cm/K. Field is applied along the c-axis for all data.

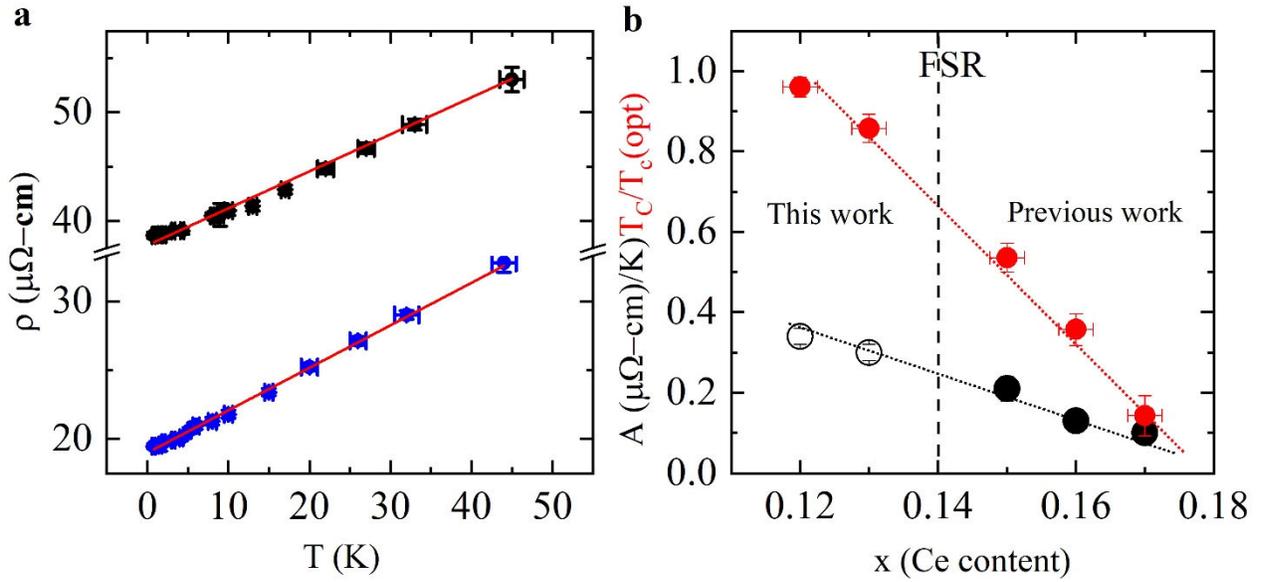

FIG. 3. (a) Resistivity vs temperature for LCCO x=0.12 at 65 T (black data points) and x=0.13 at 60 T (blue data points) from 700 mK to 45 K. The red solid line is a fit to $\rho_{xx}(T,H) = \rho_{xx}(0,H) + A(x)T$ (b) Previous work is reported in ref. 1, 2, and 4. Open circles (from figure 3(a)) and solid black circles (taken from ref-[2]) are the slopes, ($A(x)$), of the linear-in-T resistivity. The red circles are $T_c(x)$ normalized to the $T_c$ at the optimal doping (26 K). Dotted lines are guide to the eye.

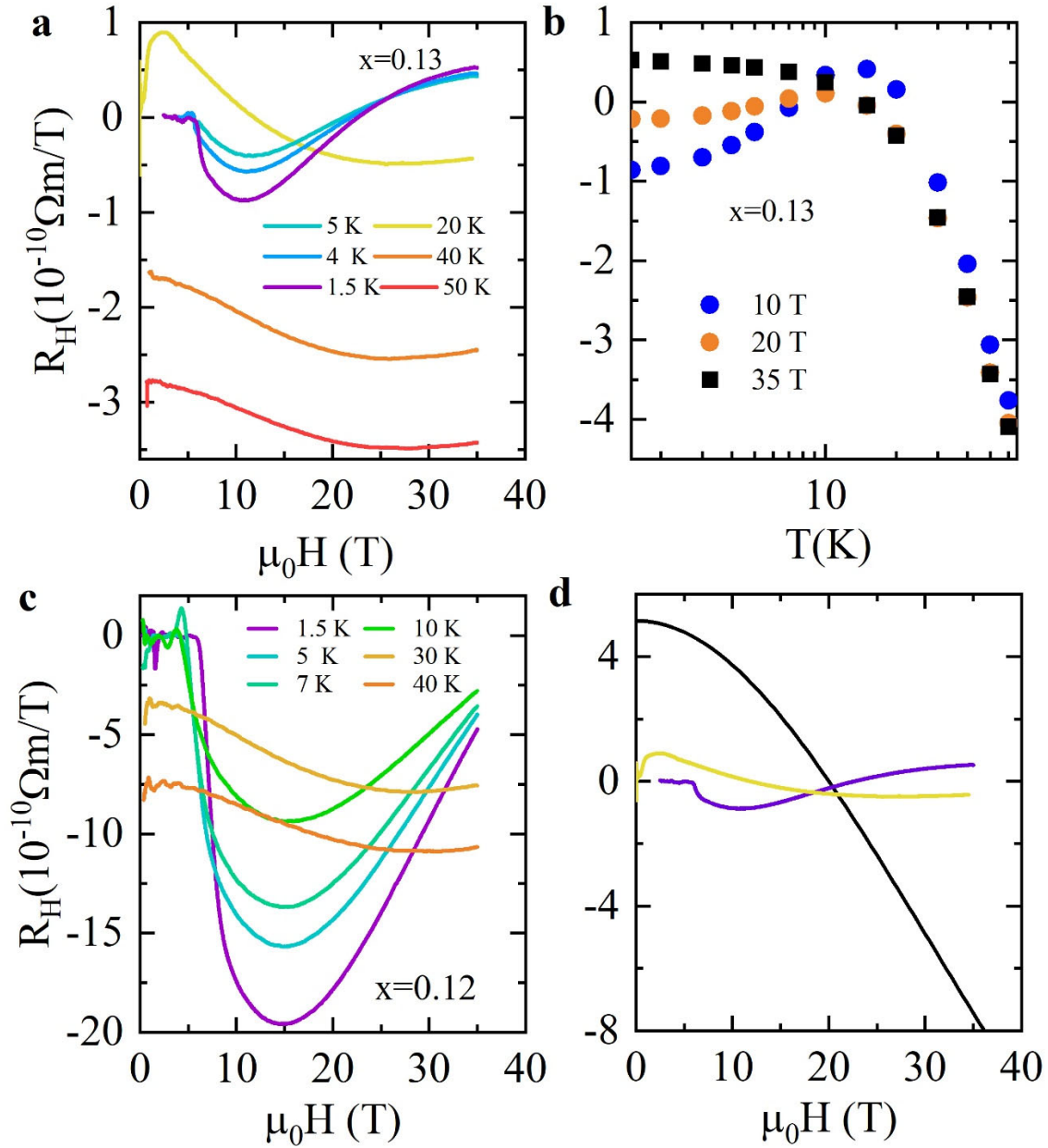

FIG. 4. (a) Hall coefficient vs field for doping x=0.13 at various temperatures. b) The Hall coefficient vs T as a function of field (data taken from Fig-4(a)). (c) Hall coefficient vs field for doping x=0.12 at various temperatures. These data are found after subtracting any magnetoresistance component by measuring $R_{xy}$ in magnetic fields from +35 T to -35 T. (d) Simulated $R_H(H)$ (black) at 2K using the estimated values of $n_e$ (1.8e27/m3), $n_h$(1e26/m3), $\mu_h$(0.025T$^{-1}$) and $\mu_e$(0.005T$^{-1}$). Blue (1.5 K) and yellow (20 K) are the measured $R_H(H)$.

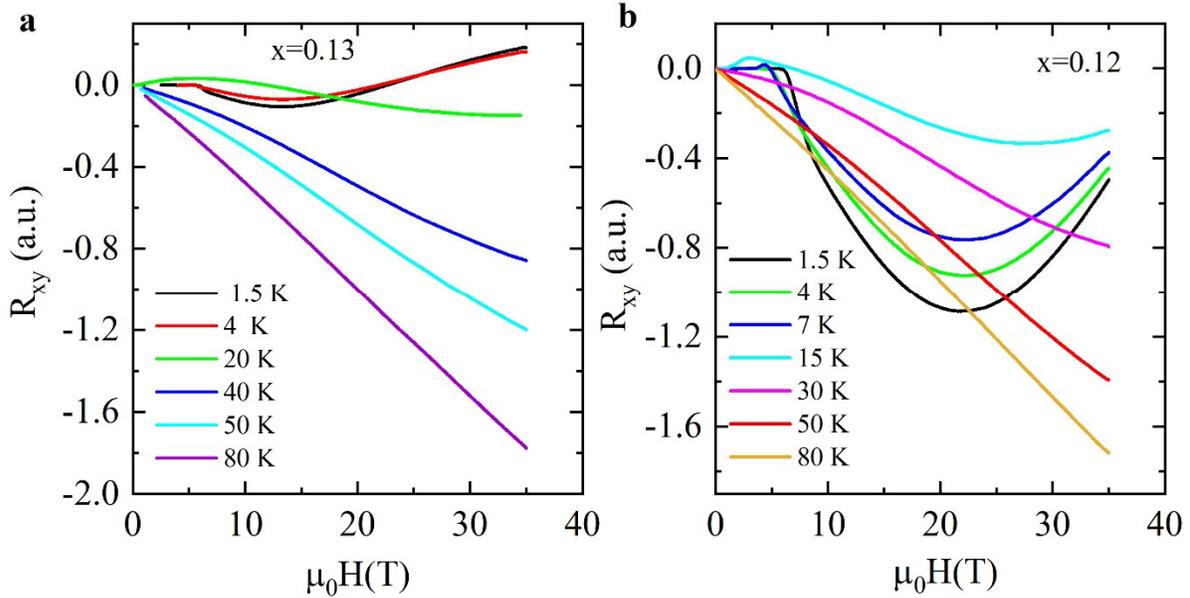

Fig.5 $R_{xy}$ vs field for doping x=0.13 (a) and x=0.12 (b) at various temperatures. These data are found after subtracting any magnetoresistance component by measuring $R_{xy}$ in magnetic fields from +35 T to -35 T.